\documentclass[a4paper,11pt]{article}
\usepackage{pos}

\usepackage{microtype}
\newcommand{\comma}{\, ,}

\newcommand{\gigaElectronVolt}{\text{GeV}}

\newcommand{\strongCoupling}{\alpha_s}
\newcommand{\couplingRescaled}{a_s}
\newcommand{\ep}{\varepsilon}
\newcommand{\mSquared}{m^2}
\newcommand{\QSquared}{Q^2}

\newcommand{\parton}{i}
\newcommand{\quark}{\text{q}}
\newcommand{\gluon}{\text{g}}
\newcommand{\vectorBoson}{\text{v}}
\newcommand{\light}{\text{l}}
\newcommand{\heavy}{\text{h}}
\newcommand{\pureSinglet}{\text{ps}}
\newcommand{\nonSinglet}{\text{ns}}
\newcommand{\flvNL}{n_\light}
\newcommand{\flvNH}{n_\heavy}
\newcommand{\flvLVVG}{\langle \light_{\{\vectorBoson,\vectorBoson\}} \rangle^{\gluon}}
\newcommand{\flvHVVG}{\langle \heavy_{\{\vectorBoson,\vectorBoson\}} \rangle^{\gluon}}

\newcommand{\colTF}{T_F}

\NewDocumentCommand{\coef}{O{N} m m}{C^{#1}_{#2,#3}}
\NewDocumentCommand{\ome}{O{N} m}{A^{#1}_{#2}}
\newcommand{\harmonicProjector}[1]{\mathcal{P}_{#1}}

\newcommand{\hpl}[1]{\text{H}_{#1}}
\newcommand{\mzv}[1]{\zeta_{#1}}
\newcommand{\numerator}[3]{\text{P}_{#1,#2,#3}}
\newcommand{\order}[1]{\mathcal{O}(#1)}
\title{Heavy-Quark Production in DIS:\\ Mellin Moments for Phenomenology}
\ShortTitle{Heavy-Quark Production in DIS: Mellin Moments for Phenomenology}

\author*[a]{Marco Klann}
\author[a]{Sven-Olaf Moch}
\author[b]{Kay Sch\"onwald}

\affiliation[a]{II.~Institut f\"ur Theoretische Physik, Universit\"at Hamburg, D-22761 Hamburg, Germany}
\affiliation[b]{Theoretical Physics Department, CERN, 1211 Geneva 23, Switzerland}

\emailAdd{marco.klann@desy.de}
\emailAdd{sven-olaf.moch@desy.de}
\emailAdd{kay.schonwald@cern.ch}

\abstract{
Heavy-quark production in deep-inelastic scattering probes the gluon distribution, and the precision of current and future data requires predictions beyond next-to-leading order.
We calculate the even Mellin moments $N = 2$ to $22$ of the next-to-leading-order heavy-quark coefficient functions for $F_2$ and $F_L$, retaining the full heavy-quark mass dependence. 
This includes the gluon channel, for which no analytic result is available.
They provide a test of the parametrisations used in phenomenology: at a representative kinematic point these are accurate to better than one per cent in the gluon channel, whereas in the quark channels we find deviations of up to $20\%$ over a wide range of Mellin moments, signalling significant differences in the corresponding $x$-space distributions.
As a new result at next-to-next-to-leading order, the lowest moment $N = 2$ has been determined exactly.
}

\FullConference{33rd International Workshop on Deep Inelastic Scattering and Related Subjects (DIS2026)\\
4--8 May 2026\\
Bologna, Italy\\}

\begin{document}
\maketitle

\section{Introduction}
\label{sec:introduction}

Heavy-quark production in deep-inelastic scattering (DIS) proceeds predominantly through photon--gluon fusion and therefore probes the gluon distribution directly, down to small momentum fractions $x$.
The combined charm and bottom data from HERA~\cite{H1:2018flt} constrain the gluon parton distribution function (PDF) and, in global fits, the charm- and bottom-quark masses~\cite{Alekhin:2017kpj}.
The Electron--Ion Collider (EIC)~\cite{AbdulKhalek:2021gbh} will measure heavy-flavour production with even higher precision.
Next-to-leading-order (NLO) predictions in quantum chromodynamics no longer match this accuracy~\cite{Alekhin:2017kpj}, so that next-to-next-to-leading-order (NNLO) corrections with the full dependence on the heavy-quark mass $m$ are required.

At NLO, the heavy-quark coefficient functions are known numerically and enter phenomenological analyses through parametrisations~\cite{Riemersma:1994hv,Hekhorn:2018ywm}.
Analytic results with the full mass dependence exist for the non-singlet~\cite{Blumlein:2016xcy,Hekhorn:2018ywm} and pure-singlet~\cite{Blumlein:2019qze} channels, but not for the gluon channel, which dominates heavy-quark production.
At NNLO, the coefficient functions are known in the asymptotic limit of photon virtualities $\QSquared \gg \mSquared$~\cite{Ablinger:2025joi} and the threshold expansion.
For general kinematics, approximate results are obtained by combining this limit with threshold and high-energy information~\cite{Kawamura:2012cr,Barontini:2026drp}.

As a first step towards the NNLO coefficient functions with full mass dependence, we compute their Mellin moments from the operator product expansion of the forward Compton amplitude~\cite{Klann:2026svr}.
Mellin moments are the natural objects of the operator product expansion and of evolution in Mellin space, and any $z$-space result can be tested against them by a single integration.
At NLO, we obtain the even moments $N = 2$ to $22$ in all channels, including the gluon channel, which allows for a direct test of the parametrisations used in phenomenology.
At NNLO, all three-loop master integrals have been computed analytically, and the moment $N = 2$ has been obtained~\cite{Klann:2026phd}.

\section{Theoretical background}
\label{sec:theoreticalBackground}

\subsection{Heavy-quark structure functions}
\label{sec:structureFunctions}

For neutral-current DIS through photon exchange with virtuality $\QSquared = -q^2 > 0$, the heavy-quark structure functions $F_k$, $k = 2, L$, factorise into the PDFs $f_\parton$ and the coefficient functions $C_{k,\parton}$,
\begin{equation}
    \frac{1}{x} F_k(x, \QSquared, \mSquared) = \sum_{\parton \in \{\quark, \bar{\quark}, \gluon\}} \int_x^{z_{\max}} \frac{\mathrm{d}z}{z} \, f_\parton\Big(\frac{x}{z}, \mu_f^2\Big) \, C_{k,\parton}(z, \QSquared, \mSquared, \mu_f^2, \mu_r^2) \comma
    \label{eq:factorisation}
\end{equation}
where $z$ is the partonic momentum fraction, bounded by $z_{\max} = 1/(1 + 4\mSquared/\QSquared)$, and $\mu_f$ and $\mu_r$ are the factorisation and renormalisation scales.
No heavy-quark PDF is included, i.e.\ there is no intrinsic heavy-quark component, and both the PDFs and the strong coupling refer to $\flvNL$ light flavours.
The partonic centre-of-mass energy squared is $s = \QSquared(1/z - 1)$, so that the production threshold $s = 4\mSquared$ lies at $z = z_{\max}$~\cite{Riemersma:1994hv}.
Mellin moments turn the convolution in Eq.~\eqref{eq:factorisation} into products of moments of the PDFs and of the coefficient functions, and large moments $N$ probe the region near threshold.

\subsection{Mellin moments from the forward Compton amplitude}
\label{sec:forwardAmplitude}

By the optical theorem, the hadronic tensor is the imaginary part of the forward Compton amplitude $T_{\mu\nu}$.
Its operator product expansion at leading twist runs over the twist-two operators of spin $N$, namely the light-quark non-singlet and singlet operators and the gluon operator.
Since only light flavours are treated as partons, heavy-quark operators do not contribute.
Projecting onto $F_2$ and $F_L$ and using crossing symmetry, one finds for even $N$
\begin{equation}
    \int_0^1 \mathrm{d}x \, x^{N-2} F_k(x, \QSquared, \mSquared) = \sum_{j \in \{\nonSinglet, \quark, \gluon\}} \ome{j} \, \coef{k}{j}(\QSquared, \mSquared) \comma
    \label{eq:momentRelation}
\end{equation}
where the hadronic operator matrix elements $\ome{j}$ are the moments of the non-singlet, singlet-quark and gluon PDFs, and the $\coef{k}{j}$ are the moments of the coefficient functions.
The odd moments follow by analytic continuation.
For quarks, we distinguish the non-singlet coefficient $\coef[N,\nonSinglet]{k}{\quark}$ from the pure-singlet coefficient $\coef[N,\pureSinglet]{k}{\quark}$, in which both photons couple to a closed quark loop and which first contributes at NLO.
Since the coefficients do not depend on the external state, the hadron can be replaced by a massless parton $\parton \in \{\quark, \gluon\}$ with momentum $p$, and the $\coef{k}{j}$ follow from partonic forward amplitudes expanded in $p$.

Since all cuts of the forward amplitude contribute, these are moments of the inclusive structure functions with $\flvNL$ light and $\flvNH$ heavy flavours.
At NLO, open heavy-quark production is isolated through the flavour factors of the graphs, together with the subtraction of the heavy-quark contribution to the light-quark form factor in the non-singlet channel~\cite{Blumlein:2016xcy}.

\subsection{Renormalisation and scheme}
\label{sec:renormalisation}

The partonic amplitudes are computed in dimensional regularisation with $D = 4 - 2\ep$ and expanded in $\couplingRescaled = \strongCoupling/(4\pi)$.
The strong coupling is renormalised in the $\overline{\text{MS}}$ scheme with $\flvNL + \flvNH$ flavours and decoupled to $\flvNL$ light flavours, as in Ref.~\cite{Riemersma:1994hv}, and the heavy-quark mass is renormalised through counterterm insertions.
The partonic operator matrix elements at $p = 0$ are evaluated in a momentum-subtraction scheme, which removes massive vacuum bubbles, so that only their tree-level values remain.
The collinear singularities are removed by mass factorisation with the known anomalous dimensions, and throughout we set $\mu_r = \mu_f = Q$.

\section{Method: harmonic projection}
\label{sec:method}

The moments are extracted from the partonic forward amplitudes by harmonic projection, the method used for the moments of the massless coefficient functions~\cite{Moch:1999eb}, which we extend here to massive quarks.
Harmonic tensors $H_q^{\mu_1 \ldots \mu_N}$ of rank $N$ are symmetric and traceless tensors built from $q$ and the metric, normalised such that $q_{\mu_1} H_q^{\mu_1 \mu_2 \ldots \mu_N} = q^2 H_q^{\mu_2 \ldots \mu_N}$.
With them, the projector
\begin{equation}
    \harmonicProjector{N} = \frac{1}{2^N N!} \, H_q^{\mu_1 \ldots \mu_N} \, \frac{\partial^N}{\partial p^{\mu_1} \cdots \partial p^{\mu_N}} \bigg|_{p = 0}
    \label{eq:harmonicProjector}
\end{equation}
acts on products of parton momenta as
\begin{equation}
    \harmonicProjector{N} \big( p^{\lambda_1} \cdots p^{\lambda_M} \big) = \delta_{MN} \, 2^{-N} H_q^{\lambda_1 \ldots \lambda_N} \comma
    \label{eq:projectorAction}
\end{equation}
so that it selects the terms with exactly $N$ powers of $p$, i.e.\ the $N$-th moment in Eq.~\eqref{eq:momentRelation}.
On the operator side, applied to the renormalised partonic invariants $T_{k,\parton}$ of the forward amplitude, it gives
\begin{equation}
    \harmonicProjector{N} (T_{k,\parton}) = (-1)^N \sum_{j \in \{\nonSinglet, \quark, \gluon\}} \coef{k}{j} \, Z_{j\parton} \, \ome[N,(0)]{\parton} \comma
    \label{eq:projectedAmplitude}
\end{equation}
where the $Z_{j\parton}$ are the operator renormalisation constants and the $\ome[N,(0)]{\parton}$ are the tree-level matrix elements, which are the only ones that survive at $p = 0$ in the scheme of Sec.~\ref{sec:renormalisation}.
Mass factorisation of Eq.~\eqref{eq:projectedAmplitude}, order by order in $\couplingRescaled$ and $\ep$, yields the finite $\coef{k}{j}$.
On the graph side, the projector acts on the numerators and on the propagators that carry $p$, and a propagator with momentum $k + p$ and mass $m$ or zero is expanded as
\begin{equation}
    \frac{1}{(k + p)^2 - \mSquared} = \frac{1}{k^2 - \mSquared} \sum_{n = 0}^{\infty} \bigg( \frac{-2 p \cdot k - p^2}{k^2 - \mSquared} \bigg)^n \comma
    \label{eq:propagatorExpansion}
\end{equation}
and all terms containing $p^2$ are annihilated by the traceless $H_q$.
Since the projection sets $p = 0$, every graph reduces to a massive two-point integral with external momentum $q$, depending on the single scale $\kappa = \mSquared/\QSquared$.
Each moment is then obtained from propagator-type integrals, exact in $\kappa$ and without any phase-space integration.

Several choices keep the computation efficient.
Since the number of terms generated by the derivatives grows with the number of propagators carrying $p$, the parton momentum is routed through as few propagators as possible.
A moment of order $N$ is even or odd under a reversal of the external momenta, so that graphs related by crossing the external gluons or photons are identified up to a sign $(-1)^N$, while graphs that vanish by colour or become scaleless after the projection are discarded.
The gluon polarisations are summed with $-g_{\mu\nu}$, since the physical polarisation sum depends explicitly on $p$ and would greatly increase the number of terms, and the unphysical polarisations are compensated by external ghosts.
Finally, $F_2$ and $F_L$ are projected with the $\ep$-independent tensors $g^{\mu\nu}$ and $4x^2 p^\mu p^\nu$, which separate the moments, so that each moment requires the amplitude only at a fixed power of $p$.
Nevertheless, the cost grows factorially with $N$, as $N = 22$ requires integrals with tensor rank and dots up to 24, and the highest moments are therefore obtained from an expansion of the full forward amplitude~\cite{Klann:2026svr}.
In all cases, the gauge parameter cancels, all poles are removed by mass factorisation, and the known massless two-loop moments are reproduced.

\section{Results}
\label{sec:results}

\subsection{Exact moments}
\label{sec:exactMoments}

We have computed the even moments $N = 2$ to $22$ of $F_2$ and $F_L$ in all channels at NLO.
They are provided in the ancillary files of Ref.~\cite{Klann:2026svr}.
The master integrals of the projected two-point functions satisfy canonical differential equations.
In the variable $\lambda$, defined by $\kappa = \lambda/(1 - \lambda)^2$, their alphabet is $\{\lambda, \lambda - 1, \lambda + 1\}$, so that every moment is a combination of harmonic polylogarithms $\hpl{\vec{a}}(\lambda)$ with rational coefficients, valid for all $\QSquared$ and $\mSquared$, with $\lambda \to 0$ for $\QSquared \gg \mSquared$ and $\lambda \to 1$ for $\QSquared \ll \mSquared$.
As an example, the lowest moment of the gluon coefficient function for $F_2$ at leading order reads
\begin{equation}
    \coef[2]{2}{\gluon} = \couplingRescaled \bigg[ \colTF \flvHVVG \flvNH \bigg( \frac{4 \hpl{0} \numerator{\gluon}{2}{02}}{3 (\lambda - 1) (\lambda + 1)^3} + \frac{\numerator{\gluon}{2}{01}}{(\lambda + 1)^2} \bigg) - \colTF \flvLVVG \flvNL \bigg] + \order{\couplingRescaled^2} \comma
    \label{eq:gluonMoment}
\end{equation}
with $\numerator{\gluon}{2}{01} = -\lambda^2 + 4\lambda - 1$ and $\numerator{\gluon}{2}{02} = \lambda^4 - 3\lambda^3 + \lambda^2 - 3\lambda + 1$, where $\flvHVVG$ and $\flvLVVG$ are the flavour factors for both photons coupling to a closed heavy- or light-quark loop~\cite{Klann:2026svr}.
At NLO, the moment $\coef[2]{2}{\gluon}$ contains harmonic polylogarithms of $\lambda$ up to weight three and $\mzv{3}$, multiplied by rational functions whose denominators are powers of $\lambda$ and $\lambda \pm 1$.

For $\QSquared \gg \mSquared$, where $\lambda = \kappa - 2\kappa^2 + \order{\kappa^3}$, the moments of $F_2$ agree with those of Ref.~\cite{Bierenbaum:2007qe}.
The heavy-quark part of Eq.~\eqref{eq:gluonMoment}, i.e.\ the bracket multiplying $\colTF \flvHVVG \flvNH$, expands, for instance, as
\begin{equation}
    -1 - \frac{4}{3} \hpl{0} + \lambda \bigg( 6 + \frac{20}{3} \hpl{0} \bigg) - \lambda^2 \bigg( 12 + \frac{44}{3} \hpl{0} \bigg) + \order{\lambda^3} \comma
    \label{eq:gluonMomentExpansion}
\end{equation}
where $\hpl{0} = \ln \lambda \simeq \ln(\mSquared/\QSquared)$ is the collinear logarithm regulated by the heavy-quark mass.
The terms in $\lambda$ and $\lambda^2$ are power corrections in $\lambda \simeq \mSquared/\QSquared$, which are absent in the asymptotic results and are contained to all orders in the exact moments.

\subsection{Benchmarking the parametrisations}
\label{sec:benchmark}

\begin{figure}[t]
    \centering
    \includegraphics[width=\textwidth]{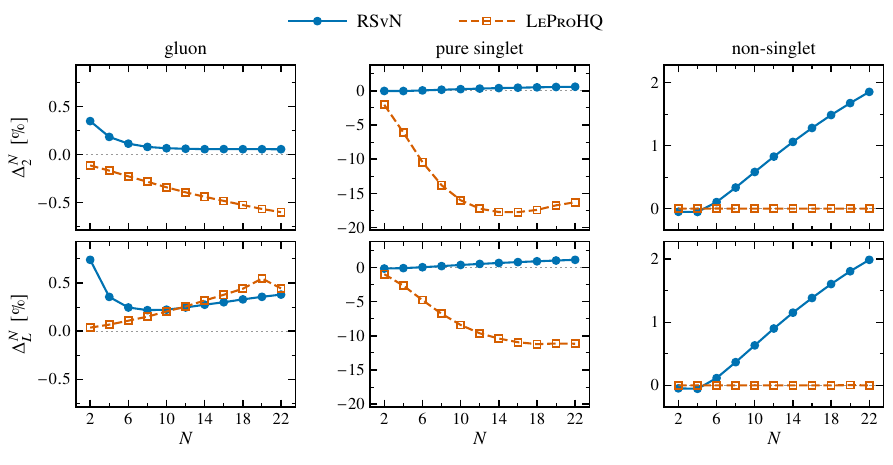}
    \caption{Relative deviation $\Delta^N_k$ of the NLO moments of the parametrisations RSvN~\cite{Riemersma:1994hv} and \textsc{LeProHQ}~\cite{Hekhorn:2018ywm} from the exact moments, for $F_2$ (top) and $F_L$ (bottom) in the gluon, pure-singlet and non-singlet channels, at $m = 2\sqrt{2}~\gigaElectronVolt$, $Q = 7~\gigaElectronVolt$ and $\mu_r = \mu_f = Q$, with the normalisation of Ref.~\cite{Klann:2026svr}.}
    \label{fig:ratioPlot}
\end{figure}

In phenomenology, the NLO coefficient functions are taken from the parametrisations of Riemersma, Smith and van Neerven (RSvN)~\cite{Riemersma:1994hv} or from \textsc{LeProHQ}~\cite{Hekhorn:2018ywm}, both given in $z$-space.
Their Mellin moments follow from a numerical integration in $z$ and can be compared directly with our exact moments (KMS), so that any difference measures the accuracy of the parametrisations.
Fig.~\ref{fig:ratioPlot} shows the relative deviation $\Delta^N_k = \coef{k}{\mathrm{param}} / \coef{k}{\mathrm{KMS}} - 1$ of the NLO terms for $m = 2\sqrt{2}~\gigaElectronVolt$ and $Q = 7~\gigaElectronVolt$, i.e.\ $\kappa = 8/49$ and $\lambda = 1/8$.
At leading order, all deviations are below $10^{-4}$.

In the gluon channel, where no analytic result in $z$ is known, both parametrisations agree with the exact moments to better than $0.8\%$ for all $N$ at this kinematic point.
The largest deviation, $0.74\%$, occurs for RSvN in $F_L$ at $N = 2$, and the deviations of RSvN decrease with $N$.
In the non-singlet channel, \textsc{LeProHQ} implements the analytic result and agrees within the precision of the numerical integration, whereas RSvN deviates by up to $2\%$ at $N = 22$.
Here, the heavy-quark contribution to the light-quark form factor is subtracted, see Sec.~\ref{sec:forwardAmplitude}.
In the pure-singlet channel, RSvN agrees within $1.2\%$, while the moments of \textsc{LeProHQ} are smaller in magnitude than the exact ones by up to $18\%$ for $F_2$ at $N = 16$ and $11\%$ for $F_L$ at $N = 18$.

\subsection{Phenomenological implications}
\label{sec:phenomenology}

The gluon channel dominates heavy-quark production, and both parametrisations describe it to better than one per cent at this kinematic point.
Since moments with $N \geq 2$ suppress the small-$z$ region, this region is tested only partially, and the comparison should be repeated for other values of $\kappa$.
For $F_L$, which the EIC will access by varying the centre-of-mass energy~\cite{AbdulKhalek:2021gbh}, the deviations in the gluon channel reach $0.74\%$ for RSvN and $0.55\%$ for \textsc{LeProHQ}.
In the quark channels, the deviations grow with $N$, indicating that the parametrisations are least accurate near threshold, $s \to 4\mSquared$.
The pure-singlet deviation of \textsc{LeProHQ} is sizeable and should be corrected in precision fits.
At NNLO, the exact moment $N = 2$~\cite{Klann:2026phd} provides a model-independent constraint on the approximate coefficient functions~\cite{Kawamura:2012cr}, which are constructed from the known limits and whose uncertainty propagates into the gluon PDF and the charm-quark mass extracted from DIS data~\cite{Alekhin:2017kpj}.

\section{Conclusions and outlook}
\label{sec:conclusions}

We have computed the even Mellin moments $N = 2$ to $22$ of all heavy-quark coefficient functions for $F_2$ and $F_L$ at NLO, exactly in $\mSquared/\QSquared$ and including the gluon channel.
They reproduce the known limit $\QSquared \gg \mSquared$ and provide a quantitative test of the parametrisations used in phenomenology.
At the kinematic point considered, the parametrisations are accurate to better than one per cent in the gluon channel, while in the quark channels RSvN deviates by up to $2\%$ and \textsc{LeProHQ} by up to $18\%$ at large $N$.

The calculation has been extended to NNLO~\cite{Klann:2026phd}.
At three loops, all master integrals, 65 unique ones for the neutral and 57 for the charged current, have been computed analytically, and the lowest moment $N = 2$ of the neutral-current coefficient functions has been determined exactly, which is the first exact NNLO result with the full mass dependence.
These moments are fully inclusive, so that final states with two and four heavy quarks are not separated.
Higher moments at NNLO are currently limited by the integration-by-parts reduction, which requires memory of the order of terabytes already for $N = 4$ and is the main computational bottleneck.

\section*{Acknowledgements}

We thank Joshua Davies and Giulio Falcioni for useful discussions and Felix Hekhorn for providing results from \textsc{LeProHQ}.
The work of M.K.\ and S.M.\ has been supported by the European Research Council through ERC Advanced Grant 101095857, \textit{Conformal-EIC}.
The research of K.S.\ was funded by the European Union's Horizon Research and Innovation Programme under the Marie Sk\l{}odowska-Curie grant agreement No.~101204018.

\bibliographystyle{JHEP}
\bibliography{bibliography}

@phdthesis{Klann:2026phd,
    author = {Klann, Marco},
    title = "{Heavy Quark Production in DIS -- Mellin Moments of Coefficient Functions at NNLO}",
    school = "Universit{\"a}t Hamburg",
    year = "2026"
}

@article{H1:2018flt,
    author = "Abramowicz, H. and others",
    collaboration = "H1, ZEUS",
    title = "{Combination and QCD analysis of charm and beauty production cross-section measurements in deep inelastic $ep$ scattering at HERA}",
    eprint = "1804.01019",
    archivePrefix = "arXiv",
    primaryClass = "hep-ex",
    reportNumber = "DESY 18-037, DESY-18-037",
    doi = "10.1140/epjc/s10052-018-5848-3",
    journal = "Eur. Phys. J. C",
    volume = "78",
    number = "6",
    pages = "473",
    year = "2018"
}

@article{Alekhin:2017kpj,
    author = {Alekhin, S. and Bl{\"u}mlein, J. and Moch, S. and Placakyte, R.},
    title = "{Parton distribution functions, $\alpha_s$, and heavy-quark masses for LHC Run II}",
    eprint = "1701.05838",
    archivePrefix = "arXiv",
    primaryClass = "hep-ph",
    reportNumber = "DESY-16-179, DO-TH-16-13",
    doi = "10.1103/PhysRevD.96.014011",
    journal = "Phys. Rev. D",
    volume = "96",
    number = "1",
    pages = "014011",
    year = "2017"
}

@article{AbdulKhalek:2021gbh,
    author = "Abdul Khalek, R. and others",
    title = "{Science Requirements and Detector Concepts for the Electron-Ion Collider}: {EIC Yellow Report}",
    eprint = "2103.05419",
    archivePrefix = "arXiv",
    primaryClass = "physics.ins-det",
    reportNumber = "BNL-220990-2021-FORE, JLAB-PHY-21-3198, LA-UR-21-20953",
    doi = "10.1016/j.nuclphysa.2022.122447",
    journal = "Nucl. Phys. A",
    volume = "1026",
    pages = "122447",
    year = "2022"
}

@article{Riemersma:1994hv,
    author = "Riemersma, S. and Smith, J. and van Neerven, W. L.",
    title = "{Rates for inclusive deep inelastic electroproduction of charm quarks at HERA}",
    eprint = "hep-ph/9411431",
    archivePrefix = "arXiv",
    reportNumber = "SMU-HEP-94-25, ITP-SB-94-59, INLO-PUB-16-94",
    doi = "10.1016/0370-2693(95)00036-K",
    journal = "Phys. Lett. B",
    volume = "347",
    pages = "143--151",
    year = "1995"
}

@article{Hekhorn:2018ywm,
    author = "Hekhorn, Felix and Stratmann, Marco",
    title = "{Next-to-Leading Order QCD Corrections to Inclusive Heavy-Flavor Production in Polarized Deep-Inelastic Scattering}",
    eprint = "1805.09026",
    archivePrefix = "arXiv",
    primaryClass = "hep-ph",
    doi = "10.1103/PhysRevD.98.014018",
    journal = "Phys. Rev. D",
    volume = "98",
    number = "1",
    pages = "014018",
    year = "2018"
}

@article{Blumlein:2016xcy,
    author = {Bl{\"u}mlein, Johannes and Falcioni, Giulio and De Freitas, Abilio},
    title = "{The Complete $O(\alpha_s^2)$ Non-Singlet Heavy Flavor Corrections to the Structure Functions $g_{1,2}^{ep}(x,Q^2)$, $F_{1,2,L}^{ep}(x,Q^2)$, $F_{1,2,3}^{\nu(\bar{\nu})}(x,Q^2)$ and the Associated Sum Rules}",
    eprint = "1605.05541",
    archivePrefix = "arXiv",
    primaryClass = "hep-ph",
    reportNumber = "DESY-15-171, DO-TH-15-14",
    doi = "10.1016/j.nuclphysb.2016.06.018",
    journal = "Nucl. Phys. B",
    volume = "910",
    pages = "568--617",
    year = "2016"
}

@article{Blumlein:2019qze,
    author = {Bl{\"u}mlein, J. and De Freitas, A. and Raab, C. G. and Sch{\"o}nwald, K.},
    title = "{The unpolarized two-loop massive pure singlet Wilson coefficients for deep-inelastic scattering}",
    eprint = "1903.06155",
    archivePrefix = "arXiv",
    primaryClass = "hep-ph",
    reportNumber = "DESY-19-038, DO-TH-18/25",
    doi = "10.1016/j.nuclphysb.2019.114659",
    journal = "Nucl. Phys. B",
    volume = "945",
    pages = "114659",
    year = "2019"
}

@article{Ablinger:2025joi,
    author = {Ablinger, J. and Behring, A. and Bl{\"u}mlein, J. and De Freitas, A. and von Manteuffel, A. and Schneider, C. and Sch{\"o}nwald, K.},
    title = "{The Single-Mass Variable Flavor Number Scheme at Three-Loop Order}",
    eprint = "2510.02175",
    archivePrefix = "arXiv",
    primaryClass = "hep-ph",
    reportNumber = "DESY 24--037, ZU-TH 58/25, RISC Report series 25-04, MPP-2025-189",
    year = "2026",
    journal = "JHEP",
    volume = "03",
    pages = "248",
    doi = "10.1007/JHEP03(2026)248",}

@article{Kawamura:2012cr,
    author = "Kawamura, H. and Lo Presti, N. A. and Moch, S. and Vogt, A.",
    title = "{On the next-to-next-to-leading order QCD corrections to heavy-quark production in deep-inelastic scattering}",
    eprint = "1205.5727",
    archivePrefix = "arXiv",
    primaryClass = "hep-ph",
    reportNumber = "KEK-TH-1378, LTH-944, DESY-12-050, LPN-12-048, SFB-CPP-12-21",
    doi = "10.1016/j.nuclphysb.2012.07.001",
    journal = "Nucl. Phys. B",
    volume = "864",
    pages = "399--468",
    year = "2012"
}

@article{Klann:2026svr,
    author = {Klann, Marco and Moch, Sven-Olaf and Sch{\"o}nwald, Kay},
    title = "{Heavy-quark production in deep-inelastic scattering -- Mellin moments of structure functions}",
    eprint = "2602.04455",
    archivePrefix = "arXiv",
    primaryClass = "hep-ph",
    reportNumber = "DESY-26-013, CERN-TH-2026-014",
    doi = "10.1016/j.nuclphysb.2026.117572",
    journal = "Nucl. Phys. B",
    volume = "1029",
    pages = "117572",
    year = "2026"
}

@article{Moch:1999eb,
    author = "Moch, S. and Vermaseren, J. A. M.",
    title = "{Deep inelastic structure functions at two loops}",
    eprint = "hep-ph/9912355",
    archivePrefix = "arXiv",
    reportNumber = "NIKHEF-99-030",
    doi = "10.1016/S0550-3213(00)00045-6",
    journal = "Nucl. Phys. B",
    volume = "573",
    pages = "853--907",
    year = "2000"
}

@article{Bierenbaum:2007qe,
     author = "Bierenbaum, Isabella and Bl{\"u}mlein, Johannes and Klein,
Sebastian",
     title = "{Two-Loop Massive Operator Matrix Elements and Unpolarized
Heavy Flavor Production at Asymptotic Values $Q^2 \gg m^2$}",
     eprint = "hep-ph/0703285",
     archivePrefix = "arXiv",
     reportNumber = "DESY-07-026, SFB-CPP-07-11",
     doi = "10.1016/j.nuclphysb.2007.04.030",
     journal = "Nucl. Phys. B",
     volume = "780",
     pages = "40--75",
     year = "2007"
}

@misc{Barontini:2026drp,
    author = "Barontini, Andrea and Bonvini, Marco and Laurenti, Niccol{\`o}",
    title = "{Implementation of DIS at N$^3$LO for PDF determination}",
    eprint = "2606.25498",
    archivePrefix = "arXiv",
    primaryClass = "hep-ph",
    month = "6",
    year = "2026"
}

\end{document}